\documentstyle[overcite,preprint,aps]{revtex}

\begin{document}

\title{Dynamics of Proton Transfer in Mesoscopic Clusters}
\author{Styliani Consta and Raymond Kapral\\
Chemical Physics Theory Group, Department of
Chemistry, \\University of Toronto, Toronto, Ontario, Canada M5S 1A1}
\date{\today}

\maketitle

\begin{abstract}
Proton transfer rates and mechanisms are studied in mesoscopic, 
liquid-state, molecular
clusters. The proton transfer occurs in a proton-ion complex solvated 
by polar molecules comprising the cluster environment. The rates and
mechanisms of the reaction are studied using both adiabatic and
non-adiabatic molecular dynamics. For large molecular clusters, the
proton-ion complex resides primarily on the surface of the cluster or 
one layer of solvent molecules inside the surface. The proton 
transfer occurs as the complex undergoes orientational fluctuations 
on the cluster surface or penetrates one solvent layer into the cluster 
leading to solvent configurations that favor the transfer. For smaller
clusters the complex resides mostly on the surface of the cluster 
and proton transfer is observed only when the complex penetrates the 
cluster and solvent configurations that favor the proton transfer are 
achieved. Quantitative information on
the cluster reaction rate constants is also presented.
\end{abstract}

\section{Introduction}
The rates and mechanisms of chemical reactions are influenced by the 
environments in which they occur. Clusters with linear dimensions
in the mesoscopic range are especially interesting reaction
environments since their properties differ from either small
molecular aggregates or bulk systems. Surface forces play
an essential role in the dynamics as do molecular fluctuations.
Since such clusters are likely involved in the reactive processes
occuring in the atmosphere, the study of their reactive dynamics 
has practical as well as fundamental interest.

A number of aspects of cluster reactions have been investigated 
recently. These investigations have examined the 
changes in the reactive dynamics that take place when a gas phase reaction is 
perturbed by clustering the reactive species with one or a few non-reactive  
or reactive molecules, reactive collisions where one or both of the 
reactive species are members of a cluster, cluster 
fragmentation reactions and ion-association reactions in water 
clusters \cite{kaukonen91}. Proton transfer barriers in small water clusters 
have been computed using density functional methods\cite{wei} and 
excited state proton transfer in large water clusters has
been investigated experimentally\cite{knochenmuss93}.
The present article continues our earlier study\cite{me94} of 
proton transfer in mesoscopic, polar molecular clusters.
In contrast to this earlier investigation which 
focused on the activation free energy and its implications for the
proton transfer reaction mechanism, here we consider the dynamics of
the transfer process. We show that, within the mesoscopic domain, the 
mechanism is a function of the cluster size and describe the distinctive 
types of solvent dynamics that induce the proton transfer process. 

The model system is essentially the same as that in our free-energy 
study and similar to that in other studies of 
proton transfer in the bulk 
phase\cite{laria92,laria94,li91,borgis92,schiffer94}. 
The system consists of a proton bound to a
pair of $A^{-}$ ions in a proton-ion complex, $(AHA)^-$. As in
Refs.~\cite{me94,laria92,laria94} the distance 
between the two $A^{-}$ ions in the complex is constrained to 
be $2.6 \AA$ so that vibration of the $A-A$ bond is not considered. 
The proton-ion intrinsic potential was constructed to model strongly
hydrogen bonded systems\cite{exp}. This intrinsic potential is given by:
\begin{equation}
 V_{int} = -\frac{1}{2}\xi_{1}u^2 + \frac{1}{4}\xi_{2}u^4 \;,
\label{vint}
\end{equation}
where $\xi_{1} = 1721.0 K/{\AA}^2$, and $\xi_{2} = 12989.0 K/{\AA}^4$.
The minima of the potential lie at $\pm 0.364\AA$
and the barrier height is about $0.2 k_{B}T$. Here $u$ is the proton
coordinate along the ion internuclear axis measured relative to the center
of the $A-A$ bond in the ion pair. In the investigation of the dynamics
of the proton transfer process the proton motion is restricted to the
one-dimensional coordinate $u$. Earlier studies of proton transfer
dynamics in the bulk phase\cite{laria94,li91} 
as well as our computations 
of the activation free energy for proton transfer in clusters\cite{me94} 
allowed for the possibility of
full three-dimensional motion of the proton. However, provided the
intrinsic potential strongly confines the proton to the vicinity of the
ion-pair internuclear axis, as was the case in these studies, the
restriction to one-dimensional motion is an accurate representation of the
dynamics. There are no fundamental difficulties in the removal of this
restriction but the computation time increases.

This complex is part of a cluster of $N$ diatomic molecules comprising 
the solvent. The diatomic solvent molecules are composed
of two interaction sites with partial charges $z_{a} = \pm 0.52e$. 
The bond length of the diatomic molecules was
fixed at $2.0\AA$ giving a molecular dipole moment of $\mu = 5.0 D$.
The interactions among solvent molecules, as well as those between the $A^{-}$
ions and the solvent, arise from site-site, 6-12 Lennard-Jones (LJ)
and Coulomb forces. The proton interacts with the solvent molecules via
Coulomb forces. The parameters for the LJ potentials are described
in \cite{laria92}. 

The dynamics of the transfer process 
\begin{equation}
(A-H \cdots A)^- \rightleftharpoons (A \cdots H-A)^-\; ,
\label{reaction}
\end{equation}
was investigated both
by treating the proton degrees of freedom adiabatically
and allowing for transitions among the protonic energy levels.
Section~\ref{adsec} describes the results obtained from the adiabatic
simulations. This section begins with an outline of the
simulation method and choice of reaction coordinate. The mechanism 
of the reaction that emerges from these calculations is then
described, followed by a calculation of the rate constant of the
reaction. Non-adiabatic dynamics is the subject of Sec.~\ref{nadsec}.
We first outline the surface hopping
method\cite{schiffer94,tully} which
was used to carry out the non-adiabatic calculations. We then 
discuss the results and
consider the modifications in the reaction rate and 
mechanism that arise as a result of non-adiabatic proton dynamics.
Section~\ref{consec} presents a summary and discussion of the results.

\section{Adiabatic Dynamics} \label{adsec}
\subsection{Simulation method}
Let $u$ be the proton
coordinate and $\{{\bf R}\}$ the set of coordinates of all other
classical particles in the system. The Schr\"{o}dinger equation for
the proton in the potential $V_p$ depending on the fixed configuration
$\{{\bf R}\}$ of the classical particles is
\begin{equation}
\begin{array}{lcl}
\hat H_{p}(\nabla_{u}, u ; \{{\bf R} \} )
\Psi_{n}(u ; \{{\bf R }\})  = 
\left[ -\displaystyle \frac{\hbar ^2}{2m_p}{\nabla_{u}}^2 + V_p(u,\{{\bf R} \})\right] 
\Psi_{n}(u ; \{{\bf R }\})    & = & \\
\epsilon_{n}( \{{\bf R} \} )\Psi_{n}(u ; \{{\bf R} \}),
\end{array}
\label{adiabhamilt}
\end{equation}
where $\hat H_{p}$ is the total proton Hamiltonian, $m_p$ is the mass of
the proton and $2\pi \hbar$ is Planck's constant. The potential energy
$V_p$ is the sum of the intrinsic (\ref{vint}) and Coulomb $V_c$ potentials, 
$V_p=V_{int}+V_c$.
The Coulomb potential $V_c$ is given by 
\begin{equation}
V_c = \sum_{i,a}\frac {z_{a}e^{2}}{\mid {\bf R}_{i,a} - {\bf R}_{CM}-
u \hat{{\bf R}}_{rel} \mid },
\end{equation}
here $a = 1,2$ labels the sites of any solvent molecule $i$, 
${\bf R}_{CM}=({\bf R}_I+{\bf R}_{II})/2$ is the center of mass of the ion
pair, $\hat{{\bf R}}_{rel}=({\bf R}_I-{\bf R}_{II})/
\mid {\bf R}_I-{\bf R}_{II} \mid$ is a unit vector directed along the 
ion-pair internuclear axis where  
${\bf R}_I$ and ${\bf R}_{II}$ are the positions of the two $A^{-}$ ions.

Classical particles with masses $m_i$ evolve according to Newton's equations
of motion,
\begin{equation}
m_{i}{\bf \ddot {R}} _{i} = -\nabla_{{\bf R}_{i}}{V_{s}}(\{{\bf R} \}) -
\nabla_{{\bf R}_{i}}\langle \Psi_{0}(\{{\bf R} \})\mid \hat H_{p}(\{{\bf R} \})
\mid \Psi_{0}(\{{\bf R} \}) \rangle,
\label{class}
\end{equation}
where the first term is force on ${\bf R}_{i}$ due to Coulomb 
and Lennard-Jones potentials that determine
the solvent-solvent and solvent ion-pair interactions;
the second term is the Hellmann-Feynman force that accounts for 
the action of the proton on the classical particles.
Here $\mid \Psi_{0}(\{{\bf R} \})\rangle$ is the ket corresponding to the
wavefunction $\Psi_{0}(u ; \{{\bf R} \})$ in the $u$ representation
and $\hat H_{p}(\{{\bf R} \})$ is the corresponding abstract 
Hamiltonian.

In order to solve (\ref{adiabhamilt}) the wave function 
$\Psi_{n}(u ; \{{\bf R} \})$ was expanded
in a linear combination of localized Gaussian functions as 
\begin{equation}
\Psi_{n}(u ; \{{\bf R} \}) = \sum_{i=1}^{n}c_{i}(\{{\bf R} \})
\phi_{i}(u)\;,
\end{equation}
with
\begin{equation}
\phi_{i}(u) = \frac{1}{{(2\pi {\sigma}_{i}^2)}^{1/4}}e^{-(u - \mu_{i})^{2}/
(4\sigma_{i}^2)}\; ,
\label{expans1}
\end{equation}
where $\mu_i$ and $\sigma_i$ denote the position and width of the Gaussian
functions. The proton-ion potential restricts the proton charge 
density to lie within the region between the two ions and it 
remains localized along the internuclear axis joining the two ions. 
The intrinsic potential intersects the interionic axis
at $\pm0.512\AA$. Eighteen Gaussian functions were used to span the 
region between the $A^-$ ions and the positions of their 
maxima were located at equally-spaced points $\mu_{i}$ between 
$-1.0\AA$ and $1.0\AA$.
The values of the widths $\sigma_{i}^2$ were taken to be
$0.0225 {\AA}^{2}$ for all basis functions. 
Use of the expansion given in (\ref{expans1}) results in a standard
(non-orthogonal) eigenvalue problem :
\begin{equation}
{\bf H c} = \epsilon {\bf S c}\; ,
\label{schrod}
\end{equation}
where ${\bf H}$ is the Hamiltonian matrix with elements
\begin{equation}
H_{ij} = \int \phi_{i}(u)H_p(\nabla_{u}, u ; \{{\bf R} \} )
\phi_{j}(u)du\; ,
\end{equation}
and ${\bf S}$ is the overlap matrix with elements
\begin{equation}
S_{ij} = \int \phi_{i}(u)\phi_{j}(u)du\; .
\end{equation}
The coefficients $c_{i}$ satisfy the normalization condition
\begin{equation}
\sum_{i,j}^{n}c_{i}S_{ij}c_{j} = 1\; .
\end{equation}
The adiabatic dynamics calculation was performed as follows :    
The Schr\"{o}dinger equation (\ref{schrod}) was solved for the ground
state wave function and energy for a given configuration of classical 
particles. Note that the
distance between the ions $A^{-}$ is kept fixed so the time-dependent
contribution to the potential arises from the positions of the 
solvent molecules in the cluster.
Then, using the ground state wave function, the Hellman-Feynman forces
given in (\ref{class}) were computed and the classical equations of motion
(\ref{class}) were integrated to yield a new classical configuration.
The Verlet algorithm\cite{verlet67} was used with time step of 
$5\times 10^{-15}s$ to integrate the classical equations. The 
constraints used to fix the intramolecular bond lengths were 
treated using the SHAKE algorithm\cite{ryckaert77}.
The constant temperature simulations were carried out using
Nos\'{e} dynamics\cite{nose84}.

\subsection{Rate constant}
One of the first steps in the study the rate of a reaction is the choice 
of a reaction coordinate. For adiabatic dynamics a natural choice for 
the proton transfer reaction coordinate is the 
expectation value of the position of the proton,
$\bar{z}_p(\{{\bf R}\})=\langle \Psi_{0}(\{{\bf R} \})\mid u
\mid \Psi_{0}(\{{\bf R} \}) \rangle$. As we shall see, this reaction
coordinate does provide a useful description of the proton transfer
reaction dynamics. However, since it depends on the expansion coefficients
of the ground state wave function which are known only numerically, it is
convenient to seek an alternative reaction coordinate whose form is known
analytically. Following earlier proton transfer studies\cite{laria92} we use as 
a reaction coordinate the solvent polarization\cite{marcus,warshel82} given by,
\begin{equation}
\Delta E(\{{\bf R}\}) = \sum_{i,a}z_{a}e
\left(\frac{1}{|{\bf R}_{i}^{a}-{\bf s}|} -
\frac{1}{|{\bf R}_{i}^{a} - {\bf s'}|}\right)\;.
\label{polariz}
\end{equation}
Here ${\bf s}$ and 
${\bf s}'$ are two points along the ion-pair axis in the reactant and
product regions. We shall show that both the expectation value of the
proton position and the solvent polarization provide equivalent
descriptions of the reaction dynamics and either is a good reaction
coordinate.

Proton transfer in clusters is an activated process\cite{me94} and 
if the free energy barrier is high enough a direct estimate of the
reaction rate will require a long molecular dynamics trajectory. In this
circumstance it is necessary estimate the reaction rate coefficient
directly from the reactive-flux correlation function\cite{yamamoto60}.
Using the polarization reaction coordinate, the (time-dependent) rate constant is
given by
\begin{equation}
k(t) = \frac{\langle \dot{\Delta E(0)}\delta(\Delta E(0)-\Delta E^{\dag})
\theta [\Delta E(t)-\Delta E^{\dag}]\rangle}
{\langle \theta [\Delta E(t)-\Delta E^{\dag}]\rangle} = k^{TST}\kappa(t)\; .
\end{equation}
where the angular brackets represent a canonical ensemble average,
$\theta (\Delta E(t))$ is the Heaviside function and $\Delta E^{\dag}$
is the value of the reaction coordinate at the barrier top 
($\Delta E^{\dag} = 0$ for our symmetrical case).
The rate constant is equal to the product of the transition 
state theory (TST) estimate of the
rate constant and the transmission coefficient $\kappa(t)$.
Using the Constrained-Reaction-Coordinate Dynamics (CRCD) ensemble\cite{carter89}
the transmission coefficient is given by,
\begin{equation}
\kappa(t) = \frac{{\langle D^{-1/2} \Delta \dot{E}\theta [\Delta E(t)]\rangle}_{c} }
{{\langle D^{-1/2} \Delta \dot{E}\theta (\Delta \dot{E} )\rangle}_{c} }\; ,
\label{transmis}
\end{equation}
where $\langle \cdots \rangle_c$ is an ensemble average in the CRCD ensemble 
where the configurational distribution is taken from the $\Delta E = 0$
constrained dynamics but the velocity distribution is that for the
system with no $\Delta E = 0$ constraint. The  
correction factor $D$ that removes the bias generated by
sampling initial configuration conditions from the constrained trajectories. For the  
polarization reaction coordinate $D$ is simply given by,
\begin{equation}
D = \frac{1}{2m}\sum_{i}{\left[\sum_{a}\nabla_{i,a}\Delta E\right]}^{2}\; ,
\end{equation}
where $m=m_i$ is the common mass of a solvent atom, 
$i$ runs over the number of molecules and $a =1, 2$. 
The TST result may be computed from the expression\cite{carter89}:
\begin{equation}
k^{TST} = {(2\pi \beta)}^{-1/2}\frac{\langle \delta (\Delta E - \Delta E^{\dag}) \rangle}
{\langle D^{-1/2} \rangle_{c} \langle \theta(\Delta E - \Delta E^{\dag})
\rangle}\; .
\label{tst}
\end{equation}

\subsection{Simulation results}
Calculations were performed for clusters of $N=20$ and $N=67$
solvent molecules at temperatures of 200$K$ and 260$K$, respectively. 
Under these conditions the clusters were in the liquid state. 
Evaporation did not occur on the time scale of the simulations,
typically several nanoseconds. 

The fact that either the average proton position or the solvent
polarization constitute acceptable reaction coordinates is demonstrated in
Fig.~\ref{rccomp} which shows that the time variations of both coordinates
track the hops of the proton between the reactant and product
configurations. Consequently we shall use both coordinates to provide
insight into the reaction mechanism and for the computation of the rate
constant.

\subsubsection{proton transfer mechanism}
The proton transfer mechanism consists of the description of the solvent 
dynamics in the course of the reaction (\ref{reaction}). In our study of
the proton transfer activation free energy we discussed the differences 
in the cluster structure when the proton was constrained to lie in the
transition state or reactant (or product) regions.\cite{me94} We may now 
describe the dynamical changes in the solvent that accomapany the transfer 
of the proton in the complex.

We begin by confirming the picture of the cluster structure that emerged
from the free energy study where a Feynman path intergral representation 
of the proton degrees of freedom was used and the
centroid\cite{gillan} of the proton 
``polymer" was taken to be the reaction coordinate.\cite{me94} We consider 
the probability density, $\rho_{\pm}(z,r)$, 
for finding a positive or negative site on a 
solvent molecule at a point $(z,r)$ in a cylindrical coordinate system
centered on the $A-A$ ion pair of the proton-ion complex with $z$ directed
along the $A-A$ axis and $r$ the radial coordinate in this cylindrical
frame. The probability density is averaged over the angle variable. Rather
than constructing this quantity by constraining the reaction coordinate to
lie in the transition state or reactant (product) regions as was the case
in Ref.~\cite{me94}, here we simply construct this quantity from a long
unconstrained adiabatic molecular dynamics run and collect statistics for
the probability density histogram only when the reaction coordinate is
found in the reactant or product configurations. The results of this
calculation are shown in Fig.~\ref{pid20} for a cluster with 20 solvent
molecules and in Fig.~\ref{pid67} for a 67-molecule cluster. The insets in
these figures show schematically the configuration of the proton-ion 
complex for the parts of the trajectory used to collect the statistics.

The structural ordering in the cluster is evident in these figures.
The cluster solvent molecules tend to strongly solvate the part of the
proton-ion complex with the more exposed negative charge; i.e., the end of
the complex which is less strongly bonded to the $H^+$ ion. This suggests
that the complex tends to ``float" on the surface of the cluster. When the
$H^+$ ion is strongly bound to one $A^-$ ion, this $A-H$ dipole has a
smaller dipole moment than that of a solvent molecule. Consequently, the
solvent-solvent interactions are stronger than the interactions between
a solvent molecule and this part of the proton-ion complex. These 
energetic arguments suggest that it may be favorable for this end of 
the proton-ion complex to reside on the surface of the cluster, a fact
borne out by our simulation results. Of course, both entropic as well as
energetic factors come into play in determining the structure of mixed
clusters\cite{mixed} but here energetic factors seem to play a dominant
role. 

There is also clear evidence of orientational order as can be seen from a
comparison of the densities for the positive and negative sites. The same
general picture applies for both the 20 and 67-molecules clusters.
However, the 67-molecules cluster is large enough to support two solvent
shells around the proton-ion complex and the presence of this second solvent
shell has some important consequences for the dynamics which we shall
describe below. The picture that emerges from these results is consistent
with that from the earlier free energy studies: the position of the proton 
in the proton-ion complex has a strong influence on the structure of the
cluster.

Insight into the reaction mechanism can be obtained by examining the
correlation between the time variation of the reaction coordinate and the
solvent-complex dynamics. Let ${\bf d}_i$ be the vector from the center
of mass of the complex to that of the $i^{th}$ solvent molecule,
${\bf d}_i = \frac{1}{2}({\bf R}_{i,1}+{\bf R}_{i,2})-{\bf R}_{CM}=
{\bf R}_{i}^{CM}-{\bf R}_{CM}$, and ${\bf d}$ the vector from the
center of mass of the complex to the center of mass of the
solvent molecules in the cluster
\begin{equation}
{\bf d} = N^{-1}\sum_{i=1}^{N}{\bf R}_{i}^{CM}-{\bf R_{CM}}=
N^{-1}\sum_{i=1}^{N}{\bf d}_{i}\;.
\end{equation}
Two quantities were used to gain insight into the nature of
the solvent-complex dynamics: $d=\mid{\bf d}\mid$ and 
$n^{\ast}  = n_{+} - n_{-} =\sum_{i=1}^{N}(\theta({\bf d}\cdot{\bf d}_i)
-\theta(-{\bf d}\cdot{\bf d}_i)) 
= -N + 2\sum_{i=1}^{N}\theta({\bf d}\cdot{\bf d}_i)$.
Here $\theta(x)$ is the Heaviside function and $n_{+}$ and $n_{-}$
are the number of solvent molecules with ${\bf d}\cdot{\bf d}_i>0$
and ${\bf d}\cdot{\bf d}_i<0$, respectively.
One expects small values of $d$ when the complex is near
the center of the cluster and large values when the complex is
on the surface. Similarly, $n^{\ast}$ varies between $n^{\ast}=0$
when the complex is symmetrically solvated in the center of the cluster
and $n^{\ast}=N$ when it lies on the surface of a symmetrical cluster.
It is possible to construct cluster configurations where the conditions
are violated. Nevertheless, these quantities provide useful indicators for 
solvent-complex dynamics and we have confirmed all aspects of our
interpretations by direct visualization of the cluster dynamics.

The results are shown in Figs.~\ref{picorr20} and \ref{picorr67}. In the
case of a 20-molecule cluster one sees that the complex resides for long
portions of time on the surface of the cluster (average distance 
$d \sim 6$) and long portions of time in the interior of the cluster
(average distance $d \sim 2$). If the complex is on the surface of the
cluster then transitions rarely occur; however, if the complex makes an
excursion into the interior (see the first large dip in $n^\ast$ and $d$ in
Fig.~\ref{picorr20}) then this excursion correlates with a proton transfer\cite{rem}. 
If the complex resides in the interior of the cluster, as is the case for times
between 2.0 and 4.0 ns in the figure, then proton transfer events 
are frequent. 

The picture is somewhat different for the 67-molecule cluster. The
proton-ion complex rarely penetrates deeply into the cluster. Some of the
proton transfer events correlate with excursions of the complex into the
cluster, but only one solvent layer deep and never far from the surface. 
However, even when the complex
``floats" on the surface of the cluster there are frequent proton transfer
events. 
Such events were not observed in the 20-molecule cluster case (although
they may occur with lower frequency on longer time scales). 
The mechanism of these transfer events becomes clear from an
examination of Fig.~\ref{pic67} which shows cluster configurations
including the complex during one proton transfer event. Initially, the
proton (black) is strongly bound to the ``white" ion in the complex. The
complex resides on the surface of the cluster with its ``gray" end
solvated in the cluster and it white end extending out of the cluster. In
the course of time a fluctuation occurs that causes the complex to assume a
configuration parallel to the surface so that there is a more nearly equal
solvation of the two ends of the complex. This is a favorable
configuration for proton transfer and transfer takes place around 
frame (d) of this figure. Once the proton transfer is
complete, then the favorable complex configuration is for the now strongly
hydrogen bonded gray end to protrude from the cluster and the weakly bound
white end to lie within the cluster.

Thus, the mechanism of the proton transfer depends on the size of the
cluster. For smaller clusters fluctuations lead to penetration of 
the complex into the interior of the cluster where proton transfer
is likely. For larger clusters 
transitions occur primarily by orientational motion of the complex on the 
surface of the cluster  or when the complex makes 
shallow penetrations into the cluster. Deep penetrations of the complex
into the cluster are rare.
For the smaller 20-molecule cluster, presumably the 
orientational motion of the complex is restricted on the surface due to
the larger surface forces. These features are borne out by an examination 
of the radial probability densities of the distance $d$, $R_c(d)$ shown in
Fig.~\ref{rhod}	(a) for a 20-molecule cluster and in (b) for a 67-molecule
cluster. For the 20-molecule cluster there are two peaks, one
corresponding to molecules in the interior of the cluster and the other for
molecules on the surface. The results for the 67-molecule cluster show a
broad single peak, with indications of some structure, 
corresponding to molecules on the surface and just
within the cluster. 

\subsubsection{rate constant}
We now turn to a quantitative treatment of the reaction rate and compute
the value of the rate constant. As described above, this calculation can
be divided into two parts: the calculation of the TST estimate of the rate
constant and the transmission coefficient. The product of these quantities
yields the full rate constant.

We begin with the computation of the TST rate constant using (\ref{tst}). 
The numerator of this expression is proportional to the probability
density of finding the reaction coordinate at the barrier top ($\Delta
E=0$) which in turn is related to the exponential of the activation free
energy. The free energy may be computed using the CRCD ensemble or
estimated directly from a long unconstrained molecular dynamics trajectory
if transitions are frequent. Figure~\ref{free} shows the free energy along
the reaction coordinate determined in this way  for a 67-molecule cluster 
along with the quadratic
approximations to the results determined from fits around the potential
minima. Note the parabolic nature of the free energy in the vicinities of
the minima; in fact, the quadratic approximation is quite accurate over
the entire range except at the barrier top where one expects the quadratic
extrapolation to overestimate the barrier height. The direct simulation 
is also most subject to error at the barrier top where the probability 
density is smallest and here one expects the direct simulation to
underestimate the barrier height. If one estimates the free energy barrier 
from the average of these two results one finds $W(0)=4.47 {\rm kT}$ for the
67-molecule cluster. Using (\ref{tst}) one finds $k^{TST}=0.04 
{\rm ps}^{-1}=1/25.0 {\rm ps}$. The corresponding results for the 20-molecule
cluster are: $W(0)=7.75 {\rm kT}$ and $k^{TST}=0.025 
{\rm ps}^{-1}=1/40.0 {\rm ps}$.

The TST rate constant may also be estimated from the analytical formula
\begin{equation}
k^{TST}= {\omega_0 \over 2 \pi} e^{-W(0)/kT}\; ,
\label{tstapprox}
\end{equation}
where $\omega_0$ is the frequency corresponding to the free energy minima. 
For comparison we may compute $k^{TST}$ using this formula 
along with the free energy determined from the proton position reaction 
coordinate instead of the solvent polarization coordinate. The results of
this computation are: $k^{TST}=0.036 {\rm ps}^{-1}=1/28.0 {\rm ps}$ for
the 67-molecule cluster and $k^{TST}=0.014 
{\rm ps}^{-1}=1/69.4 {\rm ps}$ for the 20-molecule cluster. The results
for the 67-molecule cluster are in quite good agreement and the somewhat 
poorer
results for the 20-molecule cluster arise from an inaccutate knowledge of
the barrier height.

The transmission coefficient was computed using
(\ref{transmis}). The following method was used to calculate the averages
in the CRCD ensemble: Statistically independent classical configurations 
were selected every 10 ps from
a long (1 ns in the case of 20 solvent molecule cluster and 1.24 ns 
in the case of 67 solvent molecule cluster) 
constant temperature Nos\'{e} molecular dynamics trajectory where 
the polarization reaction coordinate was constrained to lie at the
transition state ($\Delta E=0$). Initial velocities were
assigned according to the generalization of Boltzmann sampling
for rigid diatomic molecules\cite{ciccotti86}. 
For this ensemble of initial conditions, the constraint on the polarization 
coordinate was released and the trajectories were evolved forward in time 
using microcanonical molecular dynamics. This ensemble of trajectories
could then be used to generate the averages needed to compute
(\ref{transmis}). 

Figures~\ref{trans} (a) and (b), respectively, show the transmission 
coefficients as a function of time for the 67 and 20-molecule clusters. 
>From these graphs the transmission coefficients may be determined from the
plateau values and one finds $\kappa=0.4$ and $\kappa=0.5$ for the 67 and
20-molecule clusters, respectively. The results in these figures show a
rapid decay on a time scale which is less than a picosecond followed by a
somewhat longer decay, of the order of a few picoseconds, to a plateau
value. The time scale for the establishment of a plateau in $\kappa(t)$ is
longer in the cluster environment than in the bulk\cite{laria92} for a
similar but not identical intrinsic potential. This again signals
different dynamics in the cluster compared to the bulk.

The full rate constants determined from the product of $k^{TST}$ and
$\kappa$ are $k=1/62.5 {\rm ps}$ and $k=1/80.0 {\rm ps}$ for the 67 and
20-molecule clusters, respectively. Finally, we may estimate the rate
constant directly by simply monitoring the number of proton transfers in a
long unconstrained adiabatic molecular dynamics trajectory. This procedure
yields $k=1/88 {\rm ps}$ for a 67-molecule cluster where the unconstrined
trajectory had 77 proton transfer events. For a 20-molecule cluster where
the trajectory had 29 transfer events the estimated rate is 
$k=1/190 {\rm ps}$. Once again the 67-molecule results are in good
agreement. However, for the smaller 20-molecule cluster, since the free energy 
barrier is higher and the quadratic approximation is valid over a smaller range 
it is more difficult to estimate the barrier height.  In addition, proton transfer 
is a more rare event so that the direct estimate of the rate is also subject 
to uncertainties. For these reasons the rate estimates are more variable 
for the 20-molecule cluster.

\section{Non-adiabatic Dynamics} \label{nadsec}
\subsection{Simulation method}
The effect of non-adiabatic dynamics on the computation of the
rate of the reaction was taken into account by using Tully's
surface-hopping, stochastic model \cite{tully,schiffer94} 
which accounts for the possibility of quantum transitions in the
dynamics of mixed quantum-classical systems.

In this method a group of ``classical trajectories" is considered.
Each classical trajectory is evolved according to an equation of motion
similar to (\ref{class}) but with $\Psi_{0}$ replaced by $\Psi_{n}$,
any of the adiabatic functions. 
The Hamiltonian (3) characterizes the quantum system. The wave function
$\Phi(u, {\bf R}, t)$ that describes the quantum mechanical state at time
$t$, is expanded as a linear combination of the adiabatic states  
for the instantaneous ``classical configuration",
\begin{equation}
\Phi(u, \{{\bf R}\},t) = \sum_{n}C_{n}(t)\Psi_{n} (u ; \{{\bf R}\}),
\end{equation}
where $C_{n}$ are complex-valued expansion coefficients.
Substitution of the above equation into Schr\"{o}dinger equation
yields the following equation for the evolution of the expansion
coefficients:
\begin{equation}
i\hbar \dot{\tilde{C}}_k = \tilde{C}_{k}(V_{kk} - V_{00}) - i\hbar \sum_{j}
\tilde{C}_{j}{\bf \dot{R}\cdot d }_{kj} .
\label{newcoef}
\end{equation}
where
\begin{equation}
\tilde{C}_{j} = C_{j}exp(i\int_{0}^{t}dt{'}V_{00}/\hbar).
\end{equation}
Furthermore, using abstract notation
\begin{equation}
V_{kj}(\{{\bf R}\}) = \langle \Psi_i (\{{\bf R}\})\mid \hat H_p(\{{\bf R}\})
\mid \Psi_j (\{{\bf R}\}) \rangle,
\end{equation}
and the non-adiabatic coupling vector ${\bf d}_{kj}(\{{\bf R}\})$ is defined
as
\begin{equation}
{\bf d}_{kj}(\{{\bf R}\}) = \langle \Psi_k (\{{\bf R}\})\ \mid
\nabla_{\{{\bf R}\}} \mid \Psi_j (\{{\bf R}\}) \rangle.
\end{equation} 
Equations (\ref{newcoef}) are integrated simultaneously with the classical 
equations of motion. Let denote by $\Delta$ the classical time step and 
$\delta$ the 
quantum time step. At the end of each classical time step 
it is determined if a quantum transtion has taken place.
According to the ``fewest switches" algorithm the probability of 
switching from the current state $k$ to all other states $j$ during
the time interval between $t$ and $t+\Delta$ is
\begin{equation}
g_{kj} = \frac{b_{jk}(t + \Delta )\Delta }{a_{kk}(t + \Delta) },
\label{transit}
\end{equation}
where $a_{kj} = C_{k}C_{j}^{\ast}$ and 
$b_{jk} = -2 Re(a_{jk}^{\ast}{\bf \dot{R}\cdot d }_{kj} )$. 
If $g_{kj}$ is negative, it is set equal to zero.

The simulation method is described in Ref.~\cite{schiffer94}. Specifically, 
our simulations were carried out as folllows:   
Three quantum adiabatic states were used to describe the state of the
proton. Initial conditions for a  group of ``classical trajectories" 
were determined. Statistically independent classical configurations
were selected every 10 ps from a long canonical run where the polarization
reaction coordinate was constrained at zero. Initial velocities were
assigned according to the generalization of Boltzmann statistics
for rigid diatomic molecules\cite{ciccotti86}. 
Initially the total population was taken to be in the ground state, so
$\tilde{C}_{0} = 1.0$. For each of the ``classical
trajectories", initially two adiabatic steps were carried out to obtain
the non-adiabatic coupling vector.
Using the wave function $\Psi_{k}$ the Hellmann-Feynman forces 
were computed. The classical equations (\ref{class}) with $\Psi_0$ 
replaced by $\Psi_k$ were integrated using the
RATTLE algorithm \cite{andersen83}
with a time step of $\Delta=1\times 10^{-2}$ps. 
When the expectation value of proton position entered the transition region,
$-0.42<\langle\Psi_{k}(t)\mid u \mid \Psi_{k}(t)\rangle<0.42$, starting from
the previous time step, the integration
step of the classical equations of motion was changed to $\Delta = 10^{-3}$ ps
and integration was continued for 150 time steps.
Equations (\ref{newcoef})
were integrated using the Runge-Kutta method with time step
$\delta = 10^{-5}$ ps. At the end of each classical time
step the switching probability was computed using (\ref{transit})
to determine if a switch occured.
If a switch occured conservation of
the energy was satisfied by redistibution of the kinetic energy. 
Between transitions the coefficients $C_{k}(t)$ evolve coherently. When 
$\langle\Psi_{k}(t-\Delta)\mid u \mid \Psi_{k}(t-\Delta)\rangle 
<0.59 \AA <\langle\Psi_{k}(t)\mid u \mid \Psi_{k}(t)\rangle$ 
or 
$\langle\Psi_{k}(t)\mid u \mid \Psi_{k}(t)\rangle <-0.59 \AA <\langle\Psi_{k}
(t-\Delta)\mid u \mid
\Psi_{k}(t-\Delta)\rangle$, so that the proton density enters the reactant
or product regions, 
the coefficient for the current state was set equal to one, and all the
other coefficients were set equal to zero.

\subsection{simulation results}
We may now examine the effects of transitions among the protonic states 
on the proton transfer dynamics. Figure~\ref{nonadpic} shows a sample  
non-adiabatic trajectory for a 67-molecule cluster. The lower
panel in the figure is the proton position reaction coordinate 
computed using $\bar{z}_p^n=\langle \Psi_n |u | \Psi_n \rangle$, while the
upper panel shows in which of the three states the proton lies. 
We note that the polarization coordinate is also a good reaction coordinate 
in the non-adiabatic case since it tracks the proton density changes.
A number of features of these non-adiabatic trajectories are noteworthy. 

As expected,
there is a strong correlation between transitions among the protonic energy 
states and the proton transfer events. The probability of a transition 
is large when the separation between the adiabatic energy levels is small 
and the separation is smallest in the transition region. Proton transfer 
does, of course, occur in the absence of transitions to the excited protonic 
states in accord with the predictions of adiabatic dynamics but there are 
substantial modifications to the simple adiabatic model  
because the separation between the ground and first adiabatic states is comparable 
to $kT$ in the transition region. When the system is in an excited protonic 
state there is an increased probability of a proton transfer event. 
The proton density is more diffuse in the excited states 
than in the ground state. Also, when the system is in an excited protonic state 
and the reaction coordinate lies in the transition state region, the proton 
probability density is higher at extended spatial points where the proton ground state 
density is low. As a result, the proton density affects the solvent through 
Hellman-Feynman forces in ways that favour the transition state configurations.
Consequently, when the proton is in the excited states re-crossings of 
the transition state region are more frequent than when it is in the ground state.  
This effect can be seen in the figure where the number of proton transfer events, 
and attempted proton transfer events, is larger when the system is in the excited 
states than when it is in the ground state.

>From these trajectories the picture of the proton transfer process is
quite different from that for adiabatic dynamics since many proton
transfer events are accompanied by transitions to an excited protonic
state. If the system makes a transition to an excited state it
does not remain there long and quickly returns to the ground state so 
that the majority of time is spent in the ground state configuration.
>From the simulation results we estimated that the proton remains in 
excited states for only about 22 per cent of the total time. Thus, while
in the course of passage from the reactant to the product regions the
proton may make transitions into and out of an excited state, most of the
dynamics is controlled by the ground state wave function.  

For non-adiabatic dynamics, the rate constant of the reaction
was computed directly by dividing the time of the run by the
number of proton transfers between product and reactant states.
The simulation employed five trajectories (each 0.5 ns in duration) that
followed microcanonical dynamics. The average temperature of the 
system was 260K. The rate constant was estimated to be $k=0.017 ps^{-1}$
(1/59.0 ps) which is in good agreement with the adiabatic result.
The fact that the rate in the non-adiabatic case is very close to that computed 
using adiabatic dynamics most likely arises from a cancellation 
of effects: when transitions to an excited protonic state occur and there is an 
increased probability of proton transfer, there is also an increased number of 
re-crossings in the transition region which lead to a small transmission coefficient 
and reduce the rate.

\section{Conclusion} \label{consec}
Proton transfer rates and mechanisms differ significantly in the cluster
and bulk environments. As in the bulk, solvation forces play an important role in
determining the character of the cluster reaction but these sovation forces
have a distinctive character in the cluster. In the model
investigated in this paper, the $A-H$ end of the proton-ion complex, for
instance in the configuration $(A-H \cdots A)^-$, experiences interactions
with the solvent dipolar molecules that are weaker than the
solvent-solvent interactions. The $\cdots A^-$ end of the complex with the
more exposed negative charge experiences strong charge-dipole 
interactions with the solvent molecules. These specific forces are
responsible for the tendency of the complex to reside on the surface of
the cluster when the proton is in the reactant or product configurations
and for the fact that the complex tends to be oriented normal to the
surface with the $\cdots A^-$ in the cluster. 

There are strong fluctuations in these mesoscopic, liquid-state clusters.
If the cluster is small enough, as in the 20-molecule clusters studied
here, the complex resides predominantly on the surface of
the cluster and reaction occurs only when it penetrates into
the cluster.
For larger clusters, e.g. the 67-molecule cluster, 
the complex again resides predominantly on the surface or one layer of
solvent molecules inside the surface but now transfer is observed
on the cluster surface as well. 

One may consider extensions of these ideas to other situations. Whenever
there is strong charge separation in the course of reaction one might
envisage scenarios like the one described above. However, the solvation
forces could favor solvation of the complex in the interior of the
cluster. This, in turn, would possibly give rise to a transfer mechanism
similar to that in the bulk phase. The interplay between the specific
solvation forces, cluster size and the nature of the reactive species
merits further investigation.

The model for proton transfer considered here, that of a strongly hydrogen
bonded system with negligible intrinsic barrier, favors the applicability
of adiabatic dynamics. Nevertheless, we saw that although adiabatic
dynamics does provide an estimate of the rate constant,
transitions to excited states do occur and they have implications for
the mechanism. Thus, it is interesting to
investigate weakly hydrogen bonded systems where non-adiabatic effects are
even larger and could lead to a different picture of the proton transfer
process.

\acknowledgements
This work was supported in part by a grant from the Natural Sciences and
Engineering Research Council of Canada and a Killam Research Fellowship
(R.K.).

\begin{figure}[htbp]
%fig1
\caption{Comparison of the proton position, $\bar{z}_p$, and solvent
polarization, $\Delta E$ (in units of $10^{-21} C/\AA$), reaction coordinates. 
These reaction coordinates
are plotted as a function of time for a $N=20$ molecule cluster.}
\label{rccomp}
\end{figure}

\begin{figure}[htbp]
%fig2
\caption{Probability density of (a) the positive and (b) the 
negative solvent-ion sites in a cylindrical coordinate system centered 
on the proton-ion complex for a 20 molecule cluster. 
The $z$ axis is along the $A-A$ interionic axis 
and $r$ is the distance to a solvent-ion site from the interionic axis
of the $A-A$ ion pair. The probability density was constructed 
by collecting configurations every 1.25 ps while the proton charge density 
is found between $\pm 0.7 \AA$ and $\pm 1.0\AA$.} 
\label{pid20}
\end{figure}

\begin{figure}[htbp]
%fig3
\caption{Solvent ion probability densities. Same as
Fig.~\protect\ref{pid20} but for a 67-molecule cluster. 
The probability density was constructed by collecting configurations 
every 2.5 ps while the proton charge density 
is found between $\pm 0.7 \AA$ and $\pm 1.0\AA$.}
\label{pid67}
\end{figure}

\begin{figure}[htbp]
%fig4
\caption{Trajectory of the proton position, $\bar{z}_p$ (top panel),
the distance $d$ between the center of mass of the
cluster and the center of mass of the ion-pair (middle panel) and 
$n^{\ast} = n_{+} - n_{-}$ (bottom panel) as a 
function of time for a 20-molecule cluster.} 
\label{picorr20}
\end{figure}

\begin{figure}[htbp]
%fig5
\caption{Same as Fig.~\protect\ref{picorr20} but 
for a 67-molecule cluster.} 
\label{picorr67}
\end{figure}

\begin{figure}[htbp]
%fig6
\caption{Cluster configurations during a proton transfer event for a
67-molecule cluster. The configurations are separated by 2.5 ps. Time
increases from panel (a) to panel (f).}
\label{pic67}
\end{figure}

\begin{figure}[htbp]
%fig7
\caption{Radial probability density $R_c(d)= 4 \pi \rho(d) d^2$ 
versus $d$ the distance
between the center of mass of the ion pair and the center of mass of 
the solvent for (a) a 20-molecule cluster and (b) a 67-molecule cluster.
The configurations used to construct these densities were collected 
every 1.25 ps for the 20-molecule cluster and every 2.5 ps for the
67-molecule cluster in the course of 4 ns unconstrained trajectories.}
\label{rhod}
\end{figure}

\begin{figure}[htbp]
%fig8
\caption{Free energy along the polarization
coordinate for a 67-solvent molecule cluster. The 
open circles are the results from an unconstrained
trajectory of 4 ns duration. The line is computed from 
the least-squares quadratic approximation to the numerical results 
at the free energy minima.}
\label{free}
\end{figure}

\begin{figure}[htbp]
%fig9
\caption{Transmission coefficient as a function of time for (a) a 
67-solvent molecule cluster and (b) a 20-solvent molecule cluster.}
\label{trans}
\end{figure}

\begin{figure}[htbp]
%fig10
\caption{Non-adiabatic dynamics trajectory. The
upper panel shows the protonic states and the lower panel gives the 
proton position reaction coordinate as a function of time. The error 
bars represent $\pm$ one standard deviation.}
\label{nonadpic}
\end{figure}

\newpage
\renewcommand{\thepage}{}

\newpage
\begin{figure}[p]
\begin{center}
\leavevmode
\input{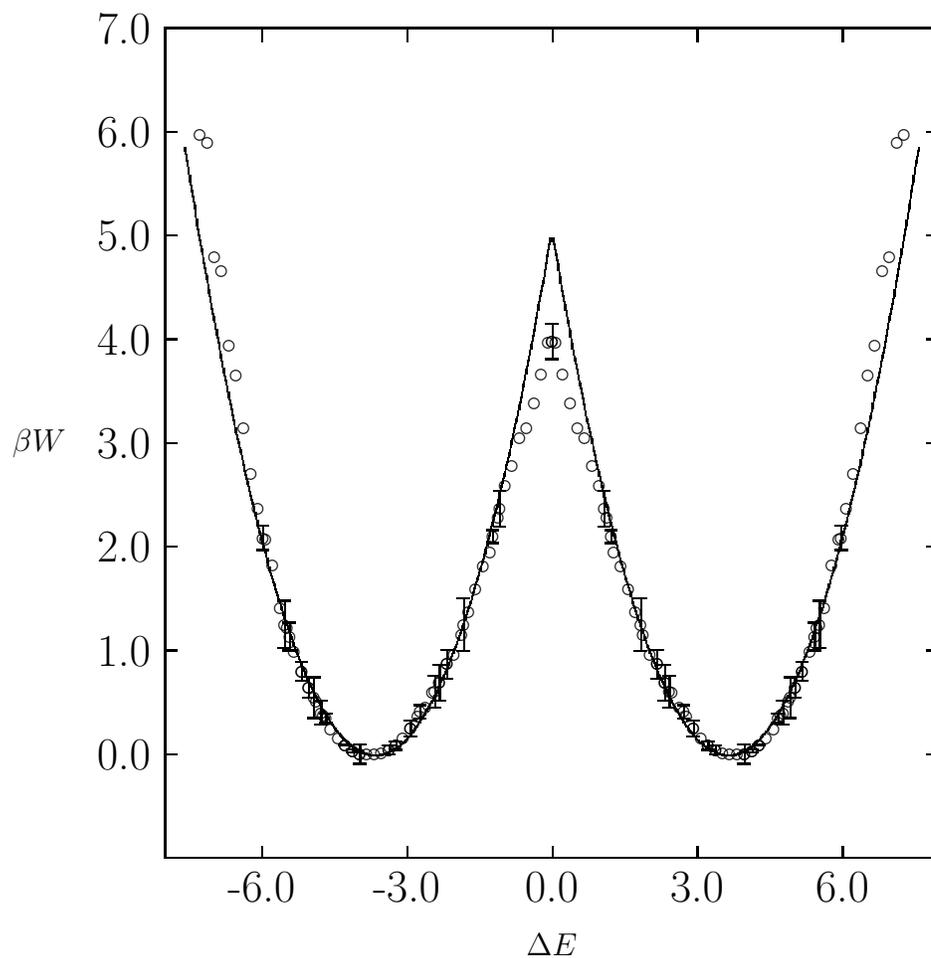}
\end{center}
\caption{Free energy along the polarization
coordinate for a 67-solvent molecule cluster. The 
open circles are the results from an unconstrained
trajectory of 4 ns duration. The line is computed from 
the least-squares quadratic approximation to the numerical results 
at the free energy minima.}
\label{fig1}
\end{figure}

\newpage
\begin{figure}[p]
\begin{center}
\leavevmode
\begin{picture}(400,500)
\put(370,330){\makebox(20,20){\Large a)}}
\put(370,30){\makebox(20,20){\Large b)}}
\put(0,300){% GNUPLOT: LaTeX picture
\setlength{\unitlength}{0.240900pt}
\ifx\plotpoint\undefined\newsavebox{\plotpoint}\fi
\sbox{\plotpoint}{\rule[-0.200pt]{0.400pt}{0.400pt}}%
\begin{picture}(1500,1080)(0,0)
\font\gnuplot=cmr10 at 10pt
\gnuplot
\sbox{\plotpoint}{\rule[-0.200pt]{0.400pt}{0.400pt}}%
\put(220.0,113.0){\rule[-0.200pt]{292.934pt}{0.400pt}}
\put(220.0,113.0){\rule[-0.200pt]{0.400pt}{227.410pt}}
\put(220.0,491.0){\rule[-0.200pt]{4.818pt}{0.400pt}}
\put(198,491){\makebox(0,0)[r]{\Large 0.2}}
\put(1416.0,491.0){\rule[-0.200pt]{4.818pt}{0.400pt}}
\put(220.0,868.0){\rule[-0.200pt]{4.818pt}{0.400pt}}
\put(198,868){\makebox(0,0)[r]{\Large 0.4}}
\put(1416.0,868.0){\rule[-0.200pt]{4.818pt}{0.400pt}}
\put(220.0,113.0){\rule[-0.200pt]{0.400pt}{4.818pt}}
\put(220,68){\makebox(0,0){\Large 0.0}}
\put(220.0,1037.0){\rule[-0.200pt]{0.400pt}{4.818pt}}
\put(828.0,113.0){\rule[-0.200pt]{0.400pt}{4.818pt}}
\put(828,68){\makebox(0,0){\Large 4.0}}
\put(828.0,1037.0){\rule[-0.200pt]{0.400pt}{4.818pt}}
\put(1436.0,113.0){\rule[-0.200pt]{0.400pt}{4.818pt}}
\put(1436,68){\makebox(0,0){\Large 8.0}}
\put(1436.0,1037.0){\rule[-0.200pt]{0.400pt}{4.818pt}}
\put(220.0,113.0){\rule[-0.200pt]{292.934pt}{0.400pt}}
\put(1436.0,113.0){\rule[-0.200pt]{0.400pt}{227.410pt}}
\put(220.0,1057.0){\rule[-0.200pt]{292.934pt}{0.400pt}}
\put(-21,585){\makebox(0,0){$\Large R_c(d)$}}
\put(828,-22){\makebox(0,0){$\Large d(\AA )$}}
\put(220.0,113.0){\rule[-0.200pt]{0.400pt}{227.410pt}}
\put(229,113){\usebox{\plotpoint}}
\multiput(430.00,113.58)(0.912,0.491){17}{\rule{0.820pt}{0.118pt}}
\multiput(430.00,112.17)(16.298,10.000){2}{\rule{0.410pt}{0.400pt}}
\multiput(448.58,123.00)(0.495,0.668){33}{\rule{0.119pt}{0.633pt}}
\multiput(447.17,123.00)(18.000,22.685){2}{\rule{0.400pt}{0.317pt}}
\multiput(466.58,147.00)(0.495,0.838){33}{\rule{0.119pt}{0.767pt}}
\multiput(465.17,147.00)(18.000,28.409){2}{\rule{0.400pt}{0.383pt}}
\multiput(484.58,177.00)(0.495,1.169){35}{\rule{0.119pt}{1.026pt}}
\multiput(483.17,177.00)(19.000,41.870){2}{\rule{0.400pt}{0.513pt}}
\multiput(503.58,221.00)(0.495,1.661){33}{\rule{0.119pt}{1.411pt}}
\multiput(502.17,221.00)(18.000,56.071){2}{\rule{0.400pt}{0.706pt}}
\multiput(521.58,280.00)(0.495,2.086){33}{\rule{0.119pt}{1.744pt}}
\multiput(520.17,280.00)(18.000,70.379){2}{\rule{0.400pt}{0.872pt}}
\multiput(539.58,354.00)(0.495,3.902){33}{\rule{0.119pt}{3.167pt}}
\multiput(538.17,354.00)(18.000,131.427){2}{\rule{0.400pt}{1.583pt}}
\multiput(557.58,492.00)(0.495,1.571){35}{\rule{0.119pt}{1.342pt}}
\multiput(556.17,492.00)(19.000,56.214){2}{\rule{0.400pt}{0.671pt}}
\multiput(576.58,546.90)(0.495,-1.122){33}{\rule{0.119pt}{0.989pt}}
\multiput(575.17,548.95)(18.000,-37.948){2}{\rule{0.400pt}{0.494pt}}
\multiput(594.58,511.00)(0.495,2.370){33}{\rule{0.119pt}{1.967pt}}
\multiput(593.17,511.00)(18.000,79.918){2}{\rule{0.400pt}{0.983pt}}
\multiput(612.58,595.00)(0.495,1.519){33}{\rule{0.119pt}{1.300pt}}
\multiput(611.17,595.00)(18.000,51.302){2}{\rule{0.400pt}{0.650pt}}
\multiput(630.58,640.02)(0.495,-2.618){35}{\rule{0.119pt}{2.163pt}}
\multiput(629.17,644.51)(19.000,-93.510){2}{\rule{0.400pt}{1.082pt}}
\multiput(649.58,545.14)(0.495,-1.661){33}{\rule{0.119pt}{1.411pt}}
\multiput(648.17,548.07)(18.000,-56.071){2}{\rule{0.400pt}{0.706pt}}
\multiput(667.58,487.90)(0.495,-1.122){33}{\rule{0.119pt}{0.989pt}}
\multiput(666.17,489.95)(18.000,-37.948){2}{\rule{0.400pt}{0.494pt}}
\multiput(685.00,450.93)(1.935,-0.477){7}{\rule{1.540pt}{0.115pt}}
\multiput(685.00,451.17)(14.804,-5.000){2}{\rule{0.770pt}{0.400pt}}
\multiput(703.58,438.89)(0.495,-2.350){35}{\rule{0.119pt}{1.953pt}}
\multiput(702.17,442.95)(19.000,-83.947){2}{\rule{0.400pt}{0.976pt}}
\multiput(722.58,359.00)(0.495,0.810){33}{\rule{0.119pt}{0.744pt}}
\multiput(721.17,359.00)(18.000,27.455){2}{\rule{0.400pt}{0.372pt}}
\multiput(740.58,379.47)(0.495,-2.483){33}{\rule{0.119pt}{2.056pt}}
\multiput(739.17,383.73)(18.000,-83.734){2}{\rule{0.400pt}{1.028pt}}
\multiput(758.58,300.00)(0.495,3.334){33}{\rule{0.119pt}{2.722pt}}
\multiput(757.17,300.00)(18.000,112.350){2}{\rule{0.400pt}{1.361pt}}
\multiput(776.00,416.92)(0.634,-0.494){27}{\rule{0.607pt}{0.119pt}}
\multiput(776.00,417.17)(17.741,-15.000){2}{\rule{0.303pt}{0.400pt}}
\put(229.0,113.0){\rule[-0.200pt]{48.421pt}{0.400pt}}
\multiput(813.58,403.00)(0.495,2.654){33}{\rule{0.119pt}{2.189pt}}
\multiput(812.17,403.00)(18.000,89.457){2}{\rule{0.400pt}{1.094pt}}
\multiput(831.58,497.00)(0.495,3.590){33}{\rule{0.119pt}{2.922pt}}
\multiput(830.17,497.00)(18.000,120.935){2}{\rule{0.400pt}{1.461pt}}
\multiput(849.00,622.93)(2.046,-0.477){7}{\rule{1.620pt}{0.115pt}}
\multiput(849.00,623.17)(15.638,-5.000){2}{\rule{0.810pt}{0.400pt}}
\multiput(868.58,619.00)(0.495,2.512){33}{\rule{0.119pt}{2.078pt}}
\multiput(867.17,619.00)(18.000,84.687){2}{\rule{0.400pt}{1.039pt}}
\multiput(886.58,708.00)(0.495,3.051){33}{\rule{0.119pt}{2.500pt}}
\multiput(885.17,708.00)(18.000,102.811){2}{\rule{0.400pt}{1.250pt}}
\multiput(904.58,816.00)(0.495,2.909){33}{\rule{0.119pt}{2.389pt}}
\multiput(903.17,816.00)(18.000,98.042){2}{\rule{0.400pt}{1.194pt}}
\multiput(922.58,914.99)(0.495,-1.093){33}{\rule{0.119pt}{0.967pt}}
\multiput(921.17,916.99)(18.000,-36.994){2}{\rule{0.400pt}{0.483pt}}
\multiput(940.58,880.00)(0.495,2.887){35}{\rule{0.119pt}{2.374pt}}
\multiput(939.17,880.00)(19.000,103.073){2}{\rule{0.400pt}{1.187pt}}
\multiput(959.58,974.02)(0.495,-4.157){33}{\rule{0.119pt}{3.367pt}}
\multiput(958.17,981.01)(18.000,-140.012){2}{\rule{0.400pt}{1.683pt}}
\multiput(977.58,841.00)(0.495,1.519){33}{\rule{0.119pt}{1.300pt}}
\multiput(976.17,841.00)(18.000,51.302){2}{\rule{0.400pt}{0.650pt}}
\multiput(995.58,871.89)(0.495,-6.965){33}{\rule{0.119pt}{5.567pt}}
\multiput(994.17,883.45)(18.000,-234.446){2}{\rule{0.400pt}{2.783pt}}
\multiput(1013.58,642.55)(0.495,-1.840){35}{\rule{0.119pt}{1.553pt}}
\multiput(1012.17,645.78)(19.000,-65.777){2}{\rule{0.400pt}{0.776pt}}
\multiput(1032.58,572.30)(0.495,-2.228){33}{\rule{0.119pt}{1.856pt}}
\multiput(1031.17,576.15)(18.000,-75.149){2}{\rule{0.400pt}{0.928pt}}
\multiput(1050.58,489.33)(0.495,-3.448){33}{\rule{0.119pt}{2.811pt}}
\multiput(1049.17,495.17)(18.000,-116.165){2}{\rule{0.400pt}{1.406pt}}
\multiput(1068.58,372.22)(0.495,-1.944){33}{\rule{0.119pt}{1.633pt}}
\multiput(1067.17,375.61)(18.000,-65.610){2}{\rule{0.400pt}{0.817pt}}
\multiput(1086.00,310.58)(0.964,0.491){17}{\rule{0.860pt}{0.118pt}}
\multiput(1086.00,309.17)(17.215,10.000){2}{\rule{0.430pt}{0.400pt}}
\multiput(1105.58,320.00)(0.495,0.810){33}{\rule{0.119pt}{0.744pt}}
\multiput(1104.17,320.00)(18.000,27.455){2}{\rule{0.400pt}{0.372pt}}
\multiput(1123.00,349.58)(0.600,0.494){27}{\rule{0.580pt}{0.119pt}}
\multiput(1123.00,348.17)(16.796,15.000){2}{\rule{0.290pt}{0.400pt}}
\multiput(1141.58,353.16)(0.495,-3.193){33}{\rule{0.119pt}{2.611pt}}
\multiput(1140.17,358.58)(18.000,-107.581){2}{\rule{0.400pt}{1.306pt}}
\multiput(1159.00,251.58)(0.680,0.494){25}{\rule{0.643pt}{0.119pt}}
\multiput(1159.00,250.17)(17.666,14.000){2}{\rule{0.321pt}{0.400pt}}
\multiput(1178.58,259.14)(0.495,-1.661){33}{\rule{0.119pt}{1.411pt}}
\multiput(1177.17,262.07)(18.000,-56.071){2}{\rule{0.400pt}{0.706pt}}
\multiput(1196.58,202.91)(0.495,-0.810){33}{\rule{0.119pt}{0.744pt}}
\multiput(1195.17,204.45)(18.000,-27.455){2}{\rule{0.400pt}{0.372pt}}
\multiput(1214.00,175.93)(1.935,-0.477){7}{\rule{1.540pt}{0.115pt}}
\multiput(1214.00,176.17)(14.804,-5.000){2}{\rule{0.770pt}{0.400pt}}
\multiput(1232.58,169.40)(0.495,-0.659){35}{\rule{0.119pt}{0.626pt}}
\multiput(1231.17,170.70)(19.000,-23.700){2}{\rule{0.400pt}{0.313pt}}
\multiput(1251.00,145.92)(0.644,-0.494){25}{\rule{0.614pt}{0.119pt}}
\multiput(1251.00,146.17)(16.725,-14.000){2}{\rule{0.307pt}{0.400pt}}
\put(795.0,403.0){\rule[-0.200pt]{4.336pt}{0.400pt}}
\multiput(1287.00,131.92)(0.600,-0.494){27}{\rule{0.580pt}{0.119pt}}
\multiput(1287.00,132.17)(16.796,-15.000){2}{\rule{0.290pt}{0.400pt}}
\put(1269.0,133.0){\rule[-0.200pt]{4.336pt}{0.400pt}}
\multiput(1324.00,116.93)(1.935,-0.477){7}{\rule{1.540pt}{0.115pt}}
\multiput(1324.00,117.17)(14.804,-5.000){2}{\rule{0.770pt}{0.400pt}}
\put(1305.0,118.0){\rule[-0.200pt]{4.577pt}{0.400pt}}
\put(1342.0,113.0){\rule[-0.200pt]{22.645pt}{0.400pt}}
\end{picture}}
\put(0,0){\input{dist67new.tex}}
\end{picture}
\end{center}
\caption{Radial probability density $R_c(d)= 4 \pi \rho(d) d^2$ 
versus $d$ the distance
between the center of mass of the ion pair and the center of mass of 
the solvent for (a) a 20-molecule cluster and (b) a 67-molecule cluster.
The configurations used to construct these densities were collected 
every 1.25 ps for the 20-molecule cluster and every 2.5 ps for the
67-molecule cluster in the course of 4 ns unconstrained trajectories.}
\label{fig3}
\end{figure}

\newpage
\begin{figure}[p]
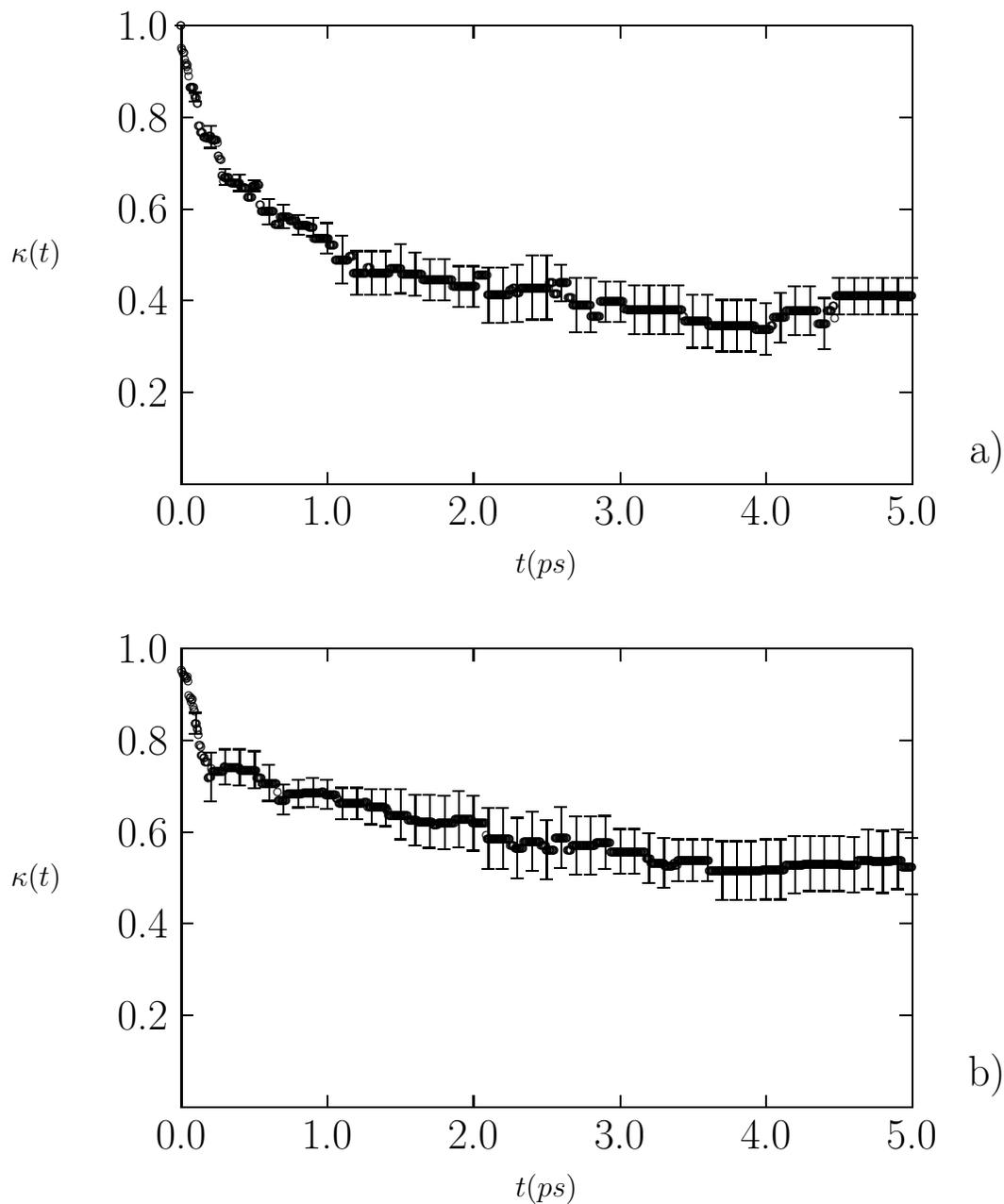

\begin{center}
\leavevmode
\begin{picture}(400,500)
\put(370,280){\makebox(20,20){\Large a)}}
\put(370,30){\makebox(20,20){\Large b)}}
\put(0,250){\input{s67T260trans.tex}}
\put(0,0){\input{s20T200trans.tex}}
\end{picture}
\end{center}
\caption{Transmission coefficient as a function of time for (a) a 
67-solvent molecule cluster and (b) a 20-solvent molecule cluster.}
\label{fig7}
\end{figure}


\begin{thebibliography}{99}
\bibitem{rem} The quantity $n^{\ast}$ shows more reliably
the location of the complex in the cluster than the
distance $d$. For instance, in Fig.~\ref{picorr20} in the vicinity of 2.6 ns 
the values of $d$ suggest that the complex is inside the cluster while the values of
$n^{\ast}$ suggest that the complex is on the surface. Apart from such infrequent
discrepancies, in our simulations the conclusions drawn from both indicators 
are generally the same.
\end{thebibliography}
\end{document}